\documentclass[aps,pra,twocolumn,balance]{revtex4} 
\usepackage{natbib}
\usepackage{epsfig}
\usepackage{graphicx}
\usepackage{amsmath}
\usepackage{amsfonts}
\usepackage{amssymb}
\usepackage{cancel}
\usepackage{graphicx}
\usepackage{ulem}
\usepackage{wrapfig}
\usepackage{epsfig}
\usepackage{graphicx}
\usepackage{amsmath}
\usepackage{amsfonts}
\usepackage{amssymb}
\usepackage{graphicx}
\usepackage{wrapfig}
\usepackage{textcomp}
\usepackage{filecontents}
\usepackage{pgf,pgfarrows,pgfnodes,pgfautomata,pgfheaps,pgfshade}
\usepackage[colorlinks=true, linkcolor=blue]{hyperref}
\usepackage{lipsum}
\usepackage{balance}
\usepackage[utf8]{inputenc}  
\usepackage{blindtext}
\usepackage{textgreek}

\begin{document}

\title{
Critical phase boundary and finite-size fluctuations  in Su-Schrieffer-Heeger  model with random inter-cell  couplings
}

\author{Dmitriy S. Shapiro$^{1,2,3}$
} \email{shapiro.dima@gmail.com}
\author{Sergey V. Remizov$^{1,4,5}$}
\author{Andrey V. Lebedev$^{1,6}$ }
\author{Danila V. Babukhin$^{1}$}
\author{Ramil S. Akzyanov$^{1,7}$}
\author{Andrey A. Zhukov$^{1}$}
\author{Leonid V. Bork$^{1}$}

	\affiliation{$^1$Dukhov Research Institute of Automatics (VNIIA),   127055 Moscow, Russia}
 \affiliation{$^2$National University of Science and Technology MISiS,   119049 Moscow, Russia}  
\affiliation{$^3$Institute for Quantum Materials and Technologies, Karlsruhe Institute of Technology, 76021 Karlsruhe, Germany}
\affiliation{$^4$V. A. Kotel'nikov Institute of Radio Engineering and Electronics, Russian Academy of Sciences, Moscow 125009, Russia}
\affiliation{$^5$Department of Physics, National Research University Higher School of Economics, Moscow 101000, Russia}
\affiliation{$^6$Moscow Institute of Physics and Technology, 141700,
Institutskii Per. 9, Dolgoprudny, Moscow Distr., Russia}
\affiliation{$^7$Institute for Theoretical and Applied Electrodynamics, Russian Academy of
Sciences, 125412 Moscow, Russia}

\begin{abstract}
A dimerized fermion chain, described by  Su-Schrieffer-Heeger (SSH) model, is a well-known example of 1D system with a non-trivial band topology.  An interplay  of disorder and topological ordering in the  SSH model is of a great  interest owing to experimental advancements in synthesized quantum simulators. 
In this work, we investigate   a special sort of a disorder when inter-cell hopping amplitudes are random.  Using a definition for    $\mathbb{Z}_2$-topological invariant $\nu\in \{ 0; 1\}$ in terms of a non-Hermitian part of the total Hamiltonian, we calculate $\langle\nu\rangle$ averaged   by random realizations. This allows to find (i) an analytical form of the critical surface that separates phases of distinct topological orders and (ii) finite size fluctuations of $\nu$ for arbitrary disorder strength. 
Numerical simulations of  the edge modes formation and  gap suppression at the transition     are  provided for finite-size system. In the end,      we discuss a  band-touching condition  derived within the averaged Green function method for a thermodynamic limit.

\end{abstract}

\maketitle

\section{Introduction}

The Su-Schrieffer-Heeger (SSH) model~\cite{PhysRevLett.42.1698, PhysRevB.22.2099} is a one-dimensional (1D) tight-binding model of a fermion chain with a certain degree of a dimerization, i.e., with alternating hopping amplitudes. It was applied  initially for a description of topological excitations, moving solitons,  in  {\it trans}-polyacetylene molecules that have a doubly degenerate ground state.
Later on, the dimerized fermion model was reexamined in condensed matter physics in contexts of topological insulators,  fractionalization of quasiparticle charge, and  adiabatic spin pump~\cite{RevModPhys.83.1057,Qi:2008aa,PhysRevLett.99.196805,PhysRevB.74.195312}. The SSH model describes a connection of geometric Zak phase and  band topology in 1D  case, where non-trivial edge modes can be formed. The experimental realizations of topological phases in SSH   become  feasible in such platforms as trapped ultracold atomic gases~\cite{Atala:2013aa,Leder:2016aa,Lohse:2016aa,xie2019topological,doi:10.1126/science.aat3406} and    superconducting   qubits~\cite{besedin2021}.    

A significant  interest is attracted by generalizations of SSH model. They  include extensions on two-chain ladders~\cite{PhysRevB.102.045108}, 2D lattice~\cite{PhysRevB.100.075437},  and long-range hopping~\cite{PhysRevB.102.205425}. This model has a deep connection    to    driven-dissipative systems described by non-Hermitian Hamiltonians~\cite{PhysRevLett.102.065703,PhysRevB.97.045106, PhysRevX.8.031079}.
As it was systematically studied in Refs.~\cite{PhysRevLett.112.206602,PhysRevB.91.085429,PhysRevLett.113.046802,PhysRevB.89.085111}, disordered versions of SSH chains reveal  transitions into topological Anderson insulator phase.  An experimental     simulation of this phenomenon in ultracold atoms was reported in Ref.~\cite{doi:10.1126/science.aat3406}.

  In this work,   we provide an analytical calculation of  $\mathbb{Z}_2$ topological index $\nu$ averaged via central limiting theorem. This solution provides a relation for the critical phase boundary. For finite size system, we provide a formula for  fluctuations of the    index and study numerically how  edge modes evolve when the disorder increases.
  
  The paper is organized as follows. We start from an  introducing of the model in Sec.~\ref{model}. In Sec.~\ref{methods} we define methods of  a calculation $\nu$ in  clean system (\ref{index-cl})  and in  disordered one (\ref{index-dis}). In Sec.~\ref{results} we present our results. In the part~\ref{averaging} the  analytic formula for  $\langle \nu\rangle$    is obtained and in~\ref{critical-ph-b} the critical phase boundary and fluctuations are calculated. Results of numerical simulations for finite-size systems are presented in Sec.~\ref{numerical}: phase diagram is analyzed in~\ref{ph-diagram} edge modes wavefunction  in~\ref{edgemodes}, and the gap suppression in~\ref{gap}.  In Sec.~\ref{summary} we conclude. The averaged Green function is found in Appendix~\ref{BA}. In Appendix~\ref{band-touching} we derive a band-touching condition from the spectral  density of states.

\section{Model}
\label{model}
\label{Hamiltonian}
The SSH Hamiltonian for a dimerized chain,
\begin{equation}
H=\sum\limits_{i=1}^N u_i \big(  a^\dagger_i b_i+   b^\dagger_i a_i \big)+w\sum\limits_{i=1}^{N-1} \big(   a^\dagger_{i+1} b_i+   b^\dagger_i a_{i+1} \big) \ , \label{H}
\end{equation}
 consists of two types of sublattices with fermion orbitals where hopping amplitudes are chosen real. The respective annihilation (creation) operators are $a_i (a^\dagger_i)$ and $b_i (b^\dagger_i)$, where $i\in \{1, \ ... , \ N\}$ and $N$ is total number of dimers.    Intra-cell hopping amplitudes (at even bonds) are constant and equal to $w$. Inter-cell   amplitudes at odd bonds, $u_i$, are  
 random  with    the average value  $\langle u_i\rangle=u$. Random deviations $\delta u_i=u_i-u$  are   uncorrelated  at different sites, i.e., $\langle\delta u_i \delta u_{i'}\rangle = \delta_{i,i'} \gamma^2$. Here, $\gamma$ is the disorder strength. 
This disorder preserves the chiral symmetry of (\ref{H}), i.e., $S H S=-H$ where $S=\prod_{i=1}^N\big(  a^\dagger_i a_i -   b^\dagger_i b_i \big)$. This symmetry indicates that zero energy modes can exist and a quantum phase transition into a topological Anderson insulator state is possible~\cite{Ryu_2010}.

The SSH model is known to have two distinct  topological phases. They are distinguished by the presence or absence of the midgap edge modes  localized at the chain ends. (The energies of these   states are exponentially close to $E=0$ in the thermodynamic limit, $N\to \infty$.) The phases have two distinct  topologies of energy bands  characterized by  $\mathbb{Z}_2$-topological invariant $\nu$ that takes two possible values, $\nu=0$   and $\nu=1$. 
There is  a spectral gap, $E_G=2|u-w|$,  in both of  these phases. It means that the topological phase  transition, which   occurs at the critical point $|w|= |u|$,   is accompanied by the  band-touching phenomenon.

 \section{Methods}
 \label{methods}
 \subsection{Topological invariant in the clean limit}
 \label{index-cl}
 In the  limit of infinite $N$ and $\gamma=0$,  the binary $\nu$ can be formulated in terms of  a geometric phase. In order to do that we rewrite the translationary invariant $H$ as an integral over   the Brillouin zone with momentum ${\mathbf{k}}\in[-\pi,\ \pi]$,
 \begin{equation}H=\int\limits_{-\pi} ^\pi\frac{d{\mathbf{k}}}{2\pi}\begin{bmatrix}a_{\mathbf{k}}^\dagger & b_{\mathbf{k}}^\dagger\end{bmatrix} \begin{bmatrix} 0 & u+w^{-i{\mathbf{k}}}\\ u+we^{i{\mathbf{k}}} & 0 \end{bmatrix} \begin{bmatrix} a_{\mathbf{k}}\\ b_{\mathbf{k}}\end{bmatrix}  \ . \label{H(k)}
 \end{equation} 
  The Fourier transform to the momentum space   reads as $a_i=\int\frac{d{\mathbf{k}}}{2\pi} a_{\mathbf{k}}e^{i{\mathbf{k}}n}$ (and similarly for $b_i$).
 The topological index  is given as $\nu=\frac{1}{\pi}\varphi_{\rm Zak}$ where $\varphi_{\rm Zak}$ is the geometric  Zak phase. It is given by an integral over the Brillouin zone  of a  Berry connection:
$
\varphi_{\rm Zak} =(-i) \int\limits_{-\pi}^\pi \langle\psi_{\mathbf{k}}|\partial_{\mathbf{k}}|\psi_{\mathbf{k}}\rangle d\mathbf{k}
$. 
Eigenfunctions $|\psi_{\mathbf{k}}\rangle $ correspond to the lower band with the dispersion $\varepsilon_{\mathbf{k}}(u,w)=-\sqrt{u^2+w^2+2uw\cos {\mathbf{k}} }$. They  read  \begin{equation}
 |\psi_{\mathbf{k}}\rangle = \frac{1}{\sqrt2}\begin{bmatrix} -e^{-i {\rm arctg} \frac{w\sin \mathbf{k}}{u+w\cos \mathbf{k}}} \\ 1 \end{bmatrix} \ . 
  \end{equation}
 Consequently, the Berry connection is a $2\pi$-periodic in $k$ function given as 
$
\langle\psi_{\mathbf{k}}|\partial_{\mathbf{k}}|\psi_{\mathbf{k}}\rangle = i\frac{w(w+u \cos k)  }{2 \varepsilon_{\mathbf{k}}^2(u,w)} 
$.
 The integral   in $\varphi_{\rm Zak}$ can be performed after the change of variables, $e^{i{\mathbf{k}}}=z$, and integration  along the contour $|z|=1$. There are two poles enclosed by the contour and their residues   contribute to $\varphi_{\rm Zak}$. The first pole is located at $z=0$ and the second one is at $z = - u/w$ if $|u/w|<1$ (or at $z = - w/u$ if $|u/w|>1$).  After some algebra one finds: 
 \begin{equation}
 \nu=\left \{ \begin{matrix} 0, \  |u|>|w|\  ;   \\ 1,   \  |u|<|w|  \ .\end{matrix}  \right. 
 \end{equation}
 If the first and last elements in the chain have hopping amplitudes $u$, then one has $\nu=0$ for $|u|> |w|$, and $\nu=1$ for $|u|< |w|$. 
 
 The value of $\nu=0$ corresponds to trivial phase with an absence of zero modes. Oppositely,  $\nu=1$ is related to a topological phase with a presence of non-trivial zero modes. 
It can be illustrated with the use of  secular equation ${\rm det} H =0$   for zero energy. It has  two complex solutions for   momenta ${\mathbf{q}}_{a,b}=\pm i \ln |w/u|+\pi$ which correspond to $a$ and $b$ sublattices, respectively. We note that in the trivial phase, $|w/u|<1$,  zero modes do not exist because wavefunctions would grow exponentially. Oppositely, in the topological phase with $|w/u|>1$ these     solutions decay and, consequently, determine wavefunctions of edge states.    One of them belongs to $a$-sublattice. It is localized at the left ($n=0$) edge and has the  exponential envelope  $|\phi^{(a)}_n\rangle \sim (-1)^{n-1}e^{-(n-1)/\xi}$.   Another edge mode is hosted by $b$-sublattice and is located at the opposite edge, $|\phi^{(b)}_n\rangle \sim (-1)^{N-n}e^{-(N-n)/\xi }$.   The  coherence length is  $\xi=(|{\rm Im}  \tilde {\mathbf{q}}_{a,b}|)^{-1}=\frac{1}{\ln |w/u|}$.  Of course, in the finite size system these solutions are not exact. As follows from symmetries of the Hamiltonian, equally weighted linear combinations of these exponential solutions can approximate exact wavefunctions of edge modes.  The overlap integral of the above solutions determines exponentially small gap between their eigenvalues.

\subsection{Topological invariant in disordered system}
 \label{index-dis}
In a system with a disorder, $\gamma\neq 0$, there is no translation invariance and the above method  can not be applied.  An alternative definition for $\nu$   is based on  an auxiliary Aharonov-Bohm phase $\Phi$  introduced   for SSH chain closed in a loop~\cite{PhysRevX.8.031079}. A periodic boundary condition (PBC) is implied in this consideration; it provides   a  gauge-invariant $\Phi$. By an analogy with the previous case of $\gamma=0$, where we dealt with  $2\pi$-periodic   Berry connection, here,    the Hamiltonian    becomes $2\pi$-periodic by $\Phi$. 
This method is based on an analysis of a complex phase of a non-Hermitian part $h$ of the total Hamiltonian represented as $H=h+h^\dagger$. 
Here, the non-Hermitian operator $h$ annihilates $b$-states and creates $a$-states only: $h=\sum\limits_{i=1}^N u_i a^\dagger_i b_i+w\sum\limits_{i=1}^{N-1}    a^\dagger_{i+1} b_i  $. In order to impose PBC one adds the term  $w a^\dagger_1 b_N$ into $h$ (and $w  b^\dagger_N a_1 $ into $h^\dagger$). The new Hamiltonian with PBC is $h_{\rm PBC}= h+w a^\dagger_1 b_N$. 
The phase can be gauged into     hopping matrix elements that is equivalent to a phase drop along the chain.   Without loss of generality,  we assume that  $\Phi$ is dropped at $N$-th bond. This transform changes a complex phase of the matrix element, i.e., one replaces $w\to w e^{i\Phi}$ in $h$. Finally,  the non-Hermitian  matrix is introduced,
 \begin{equation}\mathbf{h}_{i,j}(\Phi)=\delta_{i,j} u_i+w (\delta_{i,j+1}+e^{i\Phi}\delta_{i,j-N+1}) \ . \label{h}
\end{equation}
It parametrizes   the phase dependent Hamiltonian as $h_{\rm PBC}(\Phi)=\sum\limits_{i,j}^N a^\dagger_i \mathbf{h}_{i,j} (\Phi)b_j$.  The matrix (\ref{h}) possesses a desired $2\pi$-periodicity which provides the alternative definition of the topological invariant:
 \begin{equation}\nu=\frac{1}{2\pi i }\int\limits_0^{2\pi} d\Phi\frac{\partial}{\partial\Phi}\ln ({\rm det}\mathbf{h}_{i,j}(\Phi) )\ .\label{nu}
\end{equation}
 It can be shown that in the clean limit this definition becomes equivalent   to that formulated via the  Berry connection.

 \section{Results}
  \label{results}
 \subsection{Averaging via central limiting theorem}
  \label{averaging}
 A behavior of the invariant as a function of $w/u$ and $\gamma/u$ depends on a particular disorder realization. In what follows we study how its average by realizations, denoted as $\langle \nu\rangle$, does behave.   The average $\langle\mathcal{O}\rangle$ of a quantity $\mathcal{O}$, which depends on the set of $N$ independent random   $u_i$, is defined as $N$-dimensional integral with infinite limits
$\langle \mathcal{O} \rangle= \iint
\!\! ... \! \! \int
\mathcal{O}[u_1
,..,u_N]  \prod\limits_{i=1}^N \mathcal{P}_\gamma(u_i) du_i 
$.
Here, all of  $u_i$ are   weighted with the same probability distribution $\mathcal{P}_\gamma(\epsilon)$  normalized to the unity, $\int \mathcal{P}_\gamma(\epsilon) d\epsilon=1$. The subscript   in $\mathcal{P}_\gamma(\epsilon)$ stands for the  variance $\gamma$ provided by this distribution, i.e.,  the relation $\int  \epsilon^2 \mathcal{P}_\gamma(\epsilon)  d\epsilon- u^2=\gamma^2$ is implied. For the sake of compactness, we use  the function $p_\gamma(\epsilon )\equiv\mathcal{P}_\gamma(\epsilon -u)$ hereafter that describes random deviations  $\delta u_i$ around the mean $u$. ($p_\gamma(\epsilon )$  has the same variance as $\mathcal{P}_\gamma(\epsilon)$ but zero first moment).

According to (\ref{h}) and (\ref{nu}), we find that 
 \begin{equation}\langle \nu\rangle=\langle\theta(1-\xi) \rangle \ , \label{nu-def}
\end{equation}
where  $\theta(x)$ is the Heaviside step function, and  the random value  $\xi$ is introduced,  \begin{equation}\xi=\left(\frac{u}{w}\right)^N\prod\limits_{i=1}^N\left|1+\frac{\delta u_i}{u}\right| \ . \label{xi}
\end{equation}
As shown below, this relation between   $\xi$ and products of $u_i$ provides a non-trivial result of the averaging.

We start our consideration from the noting that the following identity holds, \begin{equation}\langle\theta(1-\xi) \rangle=\langle\theta(-\ln \xi) \rangle \ . 
\label{theta-log} 
\end{equation} 
Let us think about $\ln \xi$ as of a new random variable. According to (\ref{xi}), $\ln \xi= N\ln\frac{u}{w}+\eta$  where  $\eta=\sum\limits_{i=1}^N\ln|1+\delta u_i/u|$. The central limiting theorem can be applied at this step for $\eta$. It says that the sum of independent  and identically distributed random variables is normally distributed as  \begin{equation}P(\eta)=\frac{1}{\sqrt{2\pi}z_2}\exp\left(-\frac{1}{2 z_2^2}\left[\eta-N\left(\ln\frac{u}{w}+ z_1\right)\right]^2\right) \ .
\end{equation} 
Here, $z_1$ and $z_2$ are the first and second cumulants of $\eta$. For an arbitrary random distribution $p_\gamma(\epsilon)$, their expressions read as
\begin{equation}z_1 =\int\ln (1+\epsilon/u)p_\gamma(\epsilon)d\epsilon \label{z1}
\end{equation} 
and 
\begin{equation}
z_2 =\int\left[\ln (1+\epsilon/u)\right]^2p_\gamma(\epsilon)d\epsilon - z_1^2  \ . 
\end{equation}
Having applied the central limiting theorem, the average (\ref{theta-log}) is reduced to the integral $\langle\theta(-\ln\xi) \rangle=\int \theta(-\eta)P(\eta)  d \eta$. One obtains that $\langle\nu\rangle$   is  a   continuous   function of all Hamiltonian parameters    because the step function is smoothed after the integration by $\eta$ with  the   Gaussian profile $P(\eta)$.
The integration is performed straightforward and one arrives at 
one of central  results of this work:
\begin{equation}
\langle\nu\rangle=\frac{1}{2}\left(1-{\rm erf }
\left[\sqrt{N}\frac{\ln (u/w)+z_1( \gamma/u)}{\sqrt{2z_2(\gamma/u)}} \right]
\right) \ . \label{nu-0}
\end{equation}
This  analytical formula for  $\langle\nu\rangle$   describes a critical phase  boundary and finite-size fluctuations of the invariant near the transition. Thus, having started from the  step function with a random argument, we arrived after the averaging at the non-trivial dependence (\ref{nu-0}) which is valid for a wide range of 
  $\gamma$ and $w$.  At finite $N$ the transition between different topological phases, associated with $\langle\nu\rangle=0$ and $\langle\nu\rangle=1$, is smooth due to the averaging. In the thermodynamic limit, $N\to \infty$, there are no finite size fluctuations  and it becomes sharp.

\subsection{Critical phase boundary. Fluctuations of $\mathbb{Z}_2$ invariant}
\label{critical-ph-b}
The result (\ref{nu-0}) allows to obtain  the following   properties of the phase transition.
First, this is  the critical phase boundary. It follows from (\ref{nu-0}) under the condition $\langle\nu\rangle=1/2$. Resolving it one finds a critical $w_0$ at a given $\gamma$ and $u$. According to   (\ref{nu-0}), it reads:
 \begin{equation}
w_0=u e^{z_1(\gamma/u)
} \ . \label{gamma-0-eq}
\end{equation}
An alternative resolving of this condition with respect to $\gamma$  is complicated because one has to solve a transcendental equation. However, this can be done in the limit of weak disorder $\gamma/u\ll 1$ and weak dimerization, $u-w\ll u$, where $u>w>0$. In this limit one finds  $z_1\approx -\frac{\gamma^2}{2 u^2}$. Embedding this approximate expression into the Eq.~(\ref{gamma-0-eq}) one arrives at   the    critical disorder strength,
 \begin{equation}\gamma_0=  
 \sqrt{2 u(u-w)}   \ . \label{BA-gamma-0}
  \end{equation}   
  Similarly to the clean limit, the   topological transition driven by the disorder, which occurs at  $\gamma=\gamma_0$,  is accompanied by a  gap closing as well. The gap closing  can be shown analytically  in the limit of $u-w\ll u$ via a calculation of the density of states within first Born approximation (see Appendix~\ref{appendix}).

Second, for the binary quantity $\nu$ we immediately find that finite size fluctuations of $\Delta\nu=\nu-\langle\nu\rangle$ are  given by
\begin{equation}
\langle\Delta\nu^2\rangle= \langle\nu\rangle(1- \langle\nu\rangle) \ .  \label{nu-fluc}
\end{equation}
In the   limits of weak disorder and  dimerization mentioned above, one finds  $z_2\approx \frac{\gamma^2}{u^2}$. In this case, the fluctuations read
\begin{equation}
\langle\Delta\nu^2\rangle= \frac{1}{4}\left(1-{\rm erf}^2\left[\sqrt{\frac{N}{2}}\left(\frac{u-w}{\gamma} - \frac{\gamma}{2u}\right)\right] \right)   \ .  \label{nu-fluc-2}
\end{equation}
 The width $\Delta \gamma$  of the fluctuational region near the critical value $\gamma_0$, when other parameters are constant, is estimated as 
\begin{equation}
\Delta \gamma\sim \frac{u}{\sqrt N} \ .  \label{delta-gamma} \end{equation}
We note that only the size of the system appears in this  estimation while the dimerization parameter does not. This means that finite-size fluctuations near the critical surface are usually  not small.

\subsection{Numerical simulations}
\label{numerical}
\subsubsection{Phase diagram}
\label{ph-diagram}
\begin{figure}[h!]
	\center{\includegraphics[width=0.8\linewidth]{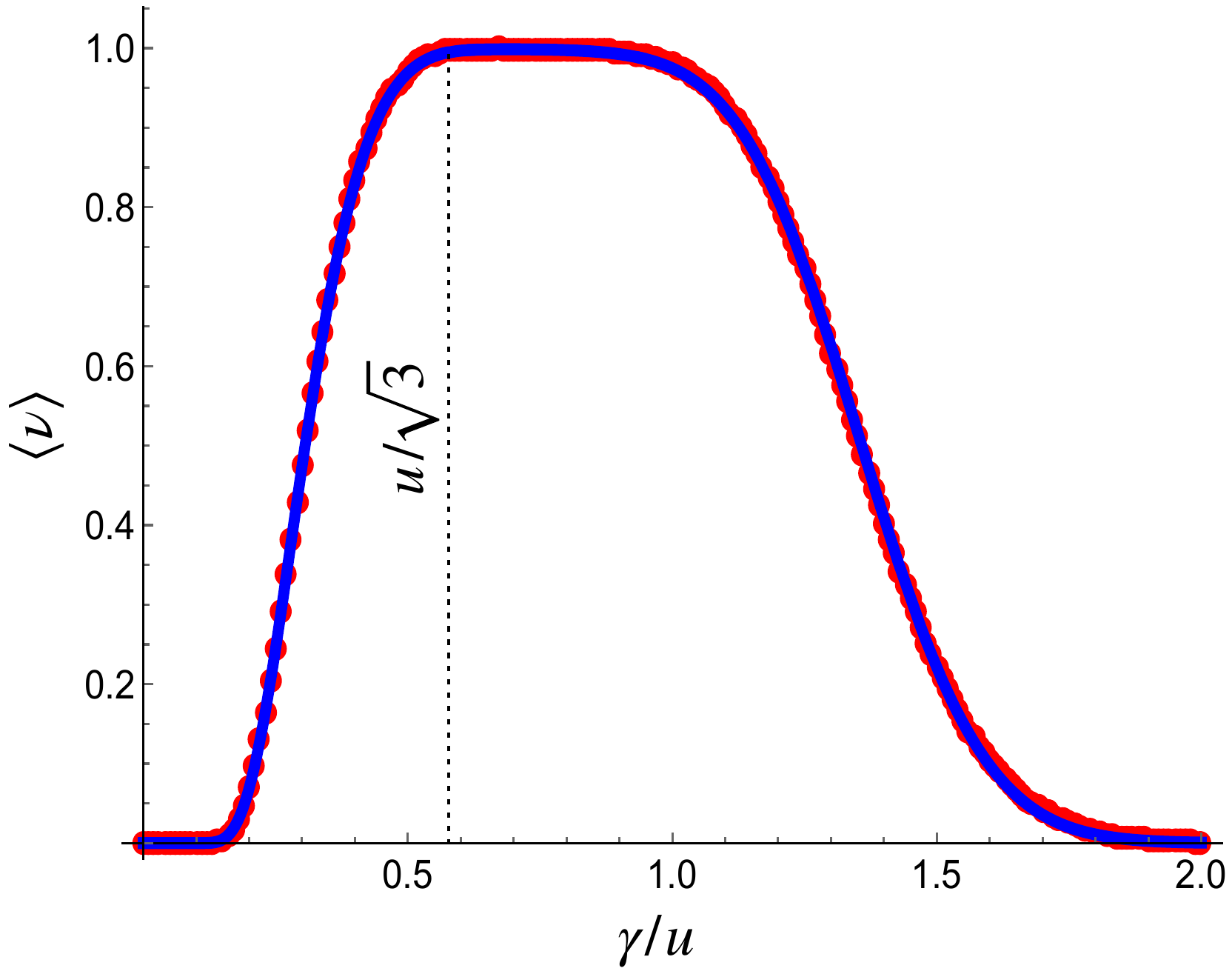}}
	\caption{ Topological   invariant  averaged by   disorder realizations $\langle\nu\rangle$ (red dots) as a function of the disorder strength, $\gamma$. The simulation is performed for the chain of $N=100$ dimers, $w/u=0.95$, and $15\times 10^3$ random distributions for $u_i$. Blue line: theoretical dependence given by Eq.~(\ref{nu-0}). The vertical dashed line is a character value of $\gamma$ that separates domains of weak and strong disorder.
}
	\label{index}
\end{figure}

The formula  (\ref{nu-0})   demonstrates a   good agreement with the data found after the numerical averaging. 
Hereafter, we assume flat distribution in $u_i\in [u-\sqrt{3}\gamma; \ u+\sqrt{3} \gamma ]$ with
 \begin{equation}
p_\gamma(\epsilon)
=\frac{1}{2\sqrt{3} \gamma} \theta(\sqrt{3}\gamma - |\epsilon|) \ .  \label{rho} 
\end{equation} 
 As demonstrated in Fig.~\ref{index}, the theoretical dependence (\ref{nu-0}) (blue curve) matches with  data of the simulation (red dots) for the chain with $N=100$ dimers. The  agreement   is observed   in a domain of  strong disorder,  $\gamma\gtrsim u$, as well. 

\begin{figure}[h]
	\center{\includegraphics[width=0.9\linewidth]{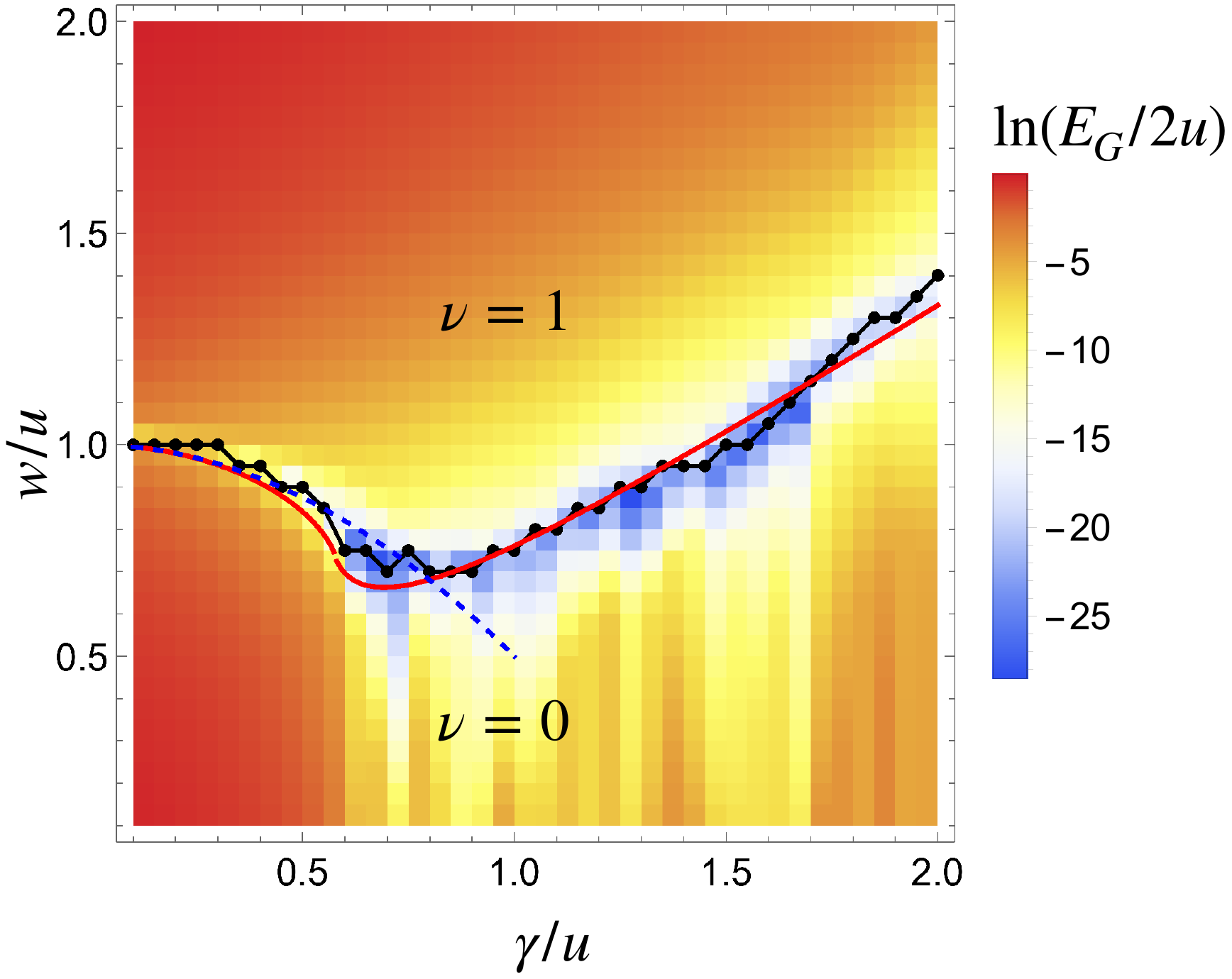}}
	\caption{ Phase diagram of the topological transition for a particular disorder realization. Density plot: the gap value in the logarithmic scale, $\ln(E_G/2u)$, as a function of $\gamma$ and $w$  in the system of $N=300$ dimers with PBC.   Black joined dots: the boundary between trivial and topological phases. Red curve: critical phase boundary   given by Eqs.~(\ref{gamma-0-eq}) and (\ref{z1-flat}). Blue dashed line: critical $w=u\sqrt{1-\frac{\gamma^2}{2u^2}}$  found for weak dimerization limit (see Eq.~\ref{BA-gamma-0}) where  the  Born approximation is justified.}
	\label{diagram}
\end{figure}

 If $w$ and $\gamma$ are varied at a constant $u$, then one arrives at a phase diagram of the disorder driven transition. 
  It is shown in the Fig.~\ref{diagram} where the gap value is plotted for the system of $N=300$ dimers  where PBC are imposed. The data shown correspond to  a particular disorder realization.  Bright red areas correspond to a finite  gap value and blue ones     to a suppressed gap. Joined black dots represent the boundary between trivial ($\nu=0$) and   topological ($\nu=1$) phases for the particular realization of $H$.  The red curve determines  the critical phase boundary  $w_0$ as a function of  $\gamma$, see Eq.~(\ref{gamma-0-eq}), where  the exponent has the following explicit form:
\begin{equation}
z_1
=- 1+\frac{u}{2\sqrt{3} \gamma} \ln \frac{u+\sqrt{3}\gamma}{|u - \sqrt{3}\gamma|} +  \frac{1}{2}\ln \left|1-\frac{3\gamma^2}{u^2}\right| \ .  \label{z1-flat} 
\end{equation}
The function is found after the integration in (\ref{z1}) with the flat distribution.

  \subsubsection{Edge modes}
  \label{edgemodes}

Here, we analyze an evolution of the edge modes wavefunction of the chain Hamiltonian (\ref{H})   when      $\gamma$ increases. In numerical simulations, we consider the eigenstates $\psi_\sigma(q E_{\rm min})$ from upper and lower bands   that have energies closest to $E=0$. Here, $\sigma=a,b$ stands for the  sublattice index and the  minimal energies    are $q E_m$ where  $E_{\rm min}={\rm min}|E_j|$ and $q=\pm 1$ due to particle-hole symmetry of $H$. For a particular realization of the disorder we calculate the  wavefunction
\begin{equation}
 |\psi|^2 =\sum\limits_{ \sigma=a,b ; q =\pm 1} |\psi_\sigma(q E_{\rm min})|^2   \label{wf-2}
\end{equation}
where a trace over    $q $- and $\sigma$-index is taken. On a next step, the averaging of (\ref{wf-2}) is performed.

\begin{figure}[h!]
	\center{\includegraphics[width=0.8\linewidth]{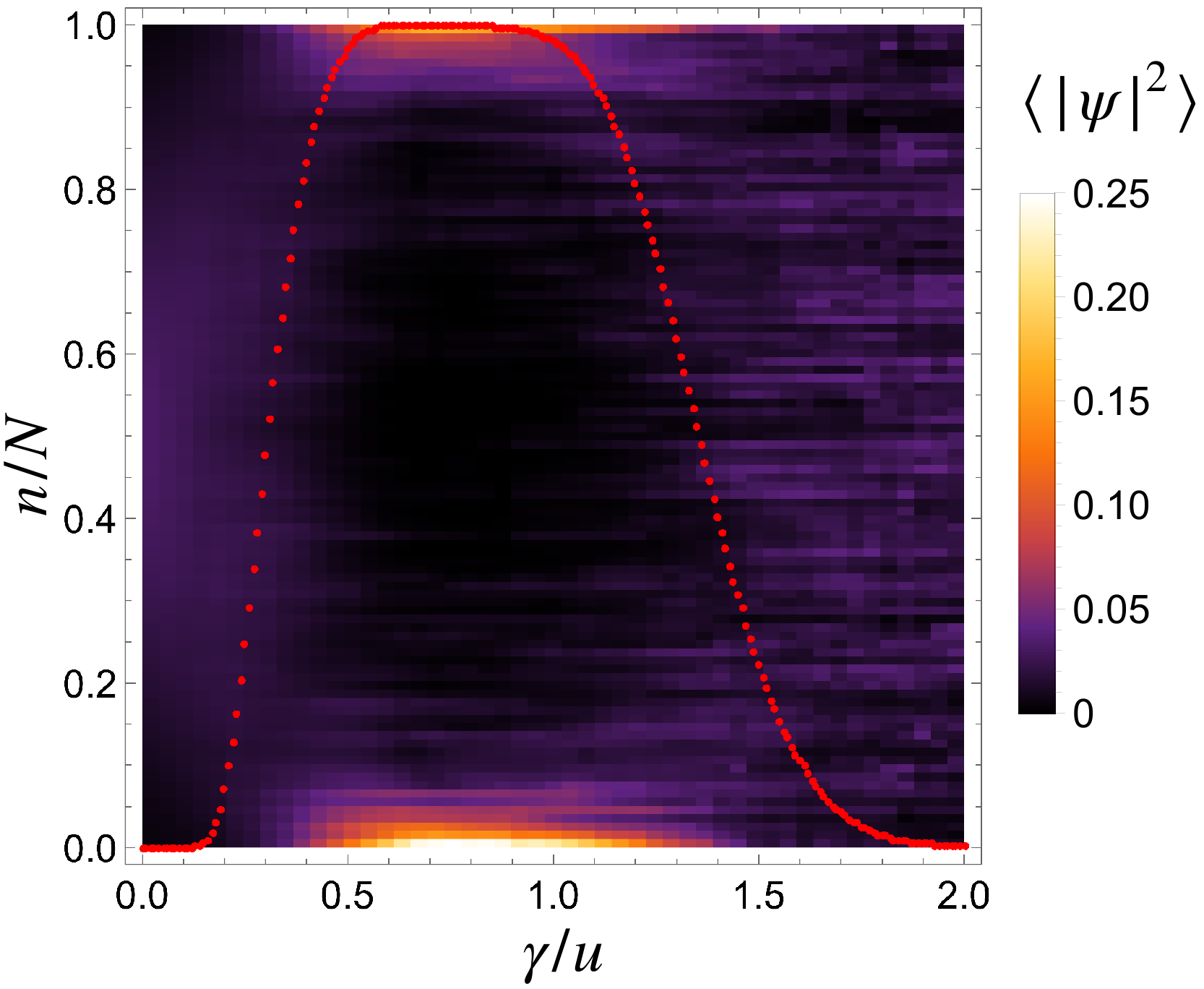}}
	\caption{ The wavefunction given by Eq.~(\ref{wf-2}) averaged by 100 realizations. Simulation is performed for the chain Hamiltonian (\ref{H}) with $N=100$ dimers and  $w/u=0.95$. Density plot for $\langle |\psi|^2\rangle$ shows a formation of edge states (red spots near $n=0$ and $n=N$) when the averaged  topological   invariant $\langle\nu\rangle$ (red dots) saturates to unity. Decrease of $\langle\nu\rangle$ when $\gamma/u\gtrsim 1$ corresponds to a disruption of the edge modes and an onset of a randomly distributed wavefunction.}
	\label{wf}
\end{figure}

In Fig.~\ref{wf} we present the data for $\langle|\psi|^2\rangle$ with the averaging  over 100 disorder realizations with $N=100$ dimers and $w/u=0.95$. 
One can see that  edge modes  appear  at $\gamma$ above $\gamma_0$ where $\langle\nu\rangle$ (red dots) saturates to the unity. Further increase of $\gamma/u\gtrsim 1$ reveals   a disruption of localized modes and a reentrance  into the non-topological phase with  the saturation of the averaged invariant to $\langle\nu\rangle=0$. The smooth decrease of $\langle\nu\rangle$ from the unity to zero means that the topological invariant is strongly fluctuating and is sensitive to a particular random realization.

 \subsubsection{Energy gap suppression}
 \label{gap}
\begin{figure}[h!]
	\center{\includegraphics[width=0.7\linewidth]{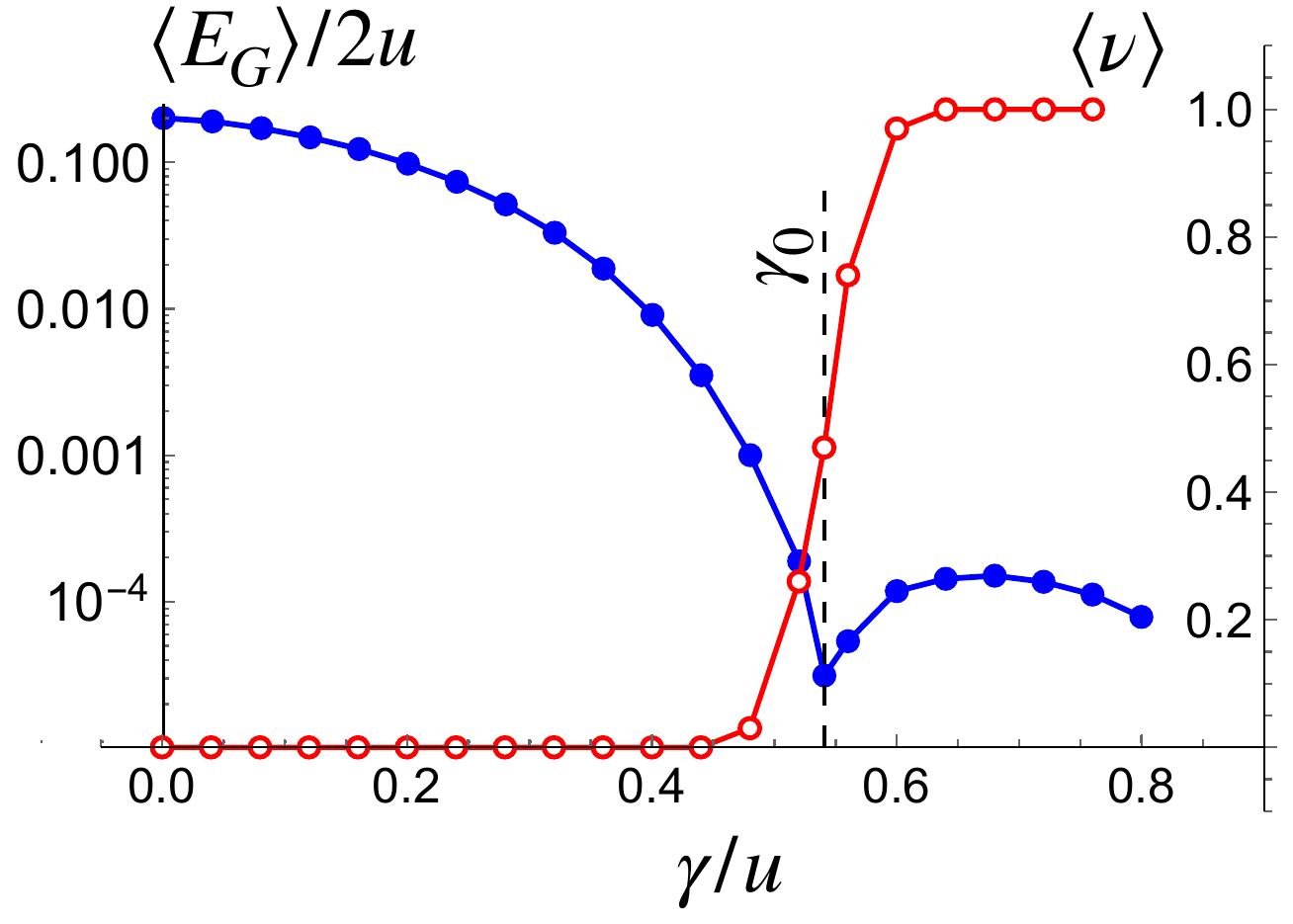}}
	\caption{
Topological phase transition driven by the disorder and suppression of the gap. Numerical data for $\langle E_G\rangle$ (blue dots, left axis) and $\langle \nu\rangle$ (red dots, right axis) averaged over 100 disorder realizations as a function of the disorder strength, $\gamma$. The simulated system involves $N=300$ dimers with PBC and $w/u=0.8$. The critical value of $\gamma_0$ corresponds to the  transition between trivial ($\nu=0$) and topological ($\nu=1$) phases and to the minumum of the gap value. }
	\label{gap-nu}
\end{figure}

  Similarly to the clean limit, the disorder driven topological transition   is accompanied by the gap closing. This effect  is described analytically  in the limit of $u-w\ll u$ via a calculation of the density of states within first Born approximation (see Appendix~\ref{appendix}). In simulations for a finite-size system  (see Fig.~\ref{gap-nu}) we observe a strong suppression of the gap,  $E_G=2E_m$, which  is decreased by five orders in the magnitude (blue dots).  The data are shown for the chain of $N=300$ dimers and averaging over 100 realizations. 
This suppression  corresponds to the band-touching phenomenon in the thermodynamic limit. At the same time, the averaged $\langle\nu\rangle$ changes smoothly from 0 to 1 (red dots) near the critical point $\gamma=\gamma_0$. The relative width of this transition, which is of the order of 10$\%$,  is in agreement with the estimation (\ref{delta-gamma}) that predicts a weak power-law decay of width with $N$.     

  \section{Discussion and outlook
}
\label{summary}

To conclude, we studied  theoretically   topological   transitions in finite-size disordered SSH model. Our findings were motivated by  state-of-the-art experiments  where    topologically ordered phases were observed in artificial dimerized chains. 
  In this work, we   derived an analytic formula for $\mathbb{Z}_2$-topological invariant $\nu$ and its fluctuations averaged by an ensemble.  The approach is based on the central limiting theorem and a non-Hermitian Hamiltonian. In particular, this  method gives an exact result for the critical surface at an arbitrary strength of the disorder.
 Our work is complementary to previous studies of   topological   phases in disordered chains~\cite{PhysRevLett.112.206602,PhysRevB.91.085429,PhysRevLett.113.046802,PhysRevB.89.085111}. 
 A particular case of    random  inter-cell tunnelling rates was considered here, however, the results  can be extended  for a more general forms of the disorder that preserve chiral symmetry.
 We also provided a detailed comparison of our findings with numerical simulations.

\begin{acknowledgments}
The reported study was supported   by Russian Foundation for Basic Research (RFBR) according to the  research project  N\textsuperscript{\underline{o}}~20-37-70028. D.S.S. also acknowledges the support  by RFBR  research project  N\textsuperscript{\underline{o}}~20-52-12034, and by DFG Grant No. MI 658/13-1 within a joint DFG-RSF project.
\end{acknowledgments}

\renewcommand\theequation{A\arabic{equation}}    
\setcounter{equation}{0}

\makeatletter
\let\@AAC@list\@empty
\makeatother

\setcounter{secnumdepth}{2}

\section{Appendix}
\label{appendix}
\subsection{Averaged Green function within Born approximation} 
\label{BA}
 Consider a retarded propagator, which has $2N\times2N$ matrix  structure in the coordinate ($n,n'$) and sublattice ($\sigma$) spaces,
 \begin{multline}[G_{n,n'}(t-t')]_{\sigma,\sigma'}= \\
 [i \delta_{n,n'}\delta_{\sigma,\sigma'}\delta(t-t')\partial_{t'} -\delta(t-t')[\mathcal{H}_{n,n'}]_{\sigma,\sigma'}]^{-1} \label{G-0}
 \end{multline} for the system with the Hamiltonian (\ref{H}). The matrix $[\mathcal{H}_{n,n'}]_{\sigma,\sigma'}=\langle n,\sigma | H |n',\sigma'\rangle$ is a projection of (\ref{H}) on the single-particle basis $|n,\sigma\rangle$ where states are defined  on $n$~-~th site and  sublattice index $\sigma=a,b$. 
 As long as the system is stationary, we use a Fourier transform by the time, i.e. $\partial_t \to - i \omega$. 
 Let us represent the Hamiltonian matrix $\mathcal{H}=\mathcal{H}^{(0)}+\mathcal{V}$, i.e. as a sum of the translational invariant part \begin{equation}\mathcal{H}^{(0)}_{n,n'}=u\delta_{n,n'} \sigma_x+w(\delta_{n,n'+1}\sigma_+ + \delta_{n+1,n'}\sigma_-) \ , 
 \end{equation} and the part with the disorder, $\mathcal{V}_{n,n'}=\epsilon_n\delta_{n,n'} \sigma_x$ that is considered as a perturbation. The sublattice indices are encoded by Pauli matrices $\sigma_x$, $\sigma_+=\frac{1}{2}(\sigma_x+i\sigma_y)$ and $\sigma_-=\frac{1}{2}(\sigma_x-i\sigma_y)$.
The Fourier transformed propagator (\ref{G-0}) is expanded in series by $\mathcal{V}$:
 \begin{equation}
 G_{n,n'}(\omega)=\mathcal{G}_{n,n'}(\omega)+\mathcal{G}_{n,k}(\omega) \sum\limits_{q=1}^{\infty}  [(\mathcal{V}  \mathcal{G}(\omega))^q]_{k,n'} \ . \label{G-1}
 \end{equation}
Here, the matrix $\mathcal{G}_{n,n'}(\omega)$ is the retarded propagator for the clean system: $\mathcal{G}_{n,n'}(\omega)=\big[\delta_{n,n'} (\omega+i\alpha) \sigma_0 -\mathcal{H}^{(0)}_{n,n'}\big]^{-1}$ where $\alpha$ is a positive infinitesimal frequency. Performing a discrete Fourier transform and using the representation of the Hamiltonian from (\ref{H(k)}), on finds $\sum\limits_{n,n'} e^{-ik n-ik'n'}\mathcal{G}_{n,n'}(\omega)=2\pi \delta({\mathbf{k}}-{\mathbf{k}}') \mathcal{G}_{\mathbf{k}}(\omega )$. Here, the bare Green function is defined for a   momentum ${\mathbf{k}}$ in  the Brillouin zone. It has a two-dimensional structure in $\sigma$-space:
   \begin{multline}
 \mathcal{G}_{\mathbf{k}}(\omega; u,w ) 
 =  \\ =\frac{(\omega+i\alpha) \sigma_0  + (u+w\cos \mathbf{k}) \sigma_x+  w  \sin \mathbf{k}  \sigma_y }{(\omega+i\alpha)^2 - \epsilon_{\mathbf{k}}^2(u,w)}  \ . \label{bare-GF}
 \end{multline}    
 Let us return back to the Green function in the coordinate space (\ref{G-1}). The averaging of  $G_{n,n'}$ by disorder realizations makes it translational invariant. Below we perform this calculation in a first order self-energy  (Born approximation) where  one takes into account only Fock-type diagram~\cite{altland2010condensed}.
 The averaged  Green function  does not include crossed line and ``rainbow''    diagrams; it is represented as follows:
  \begin{equation}
\langle G_{n,n'}\rangle  =\mathcal{G}_{n,n'} +\mathcal{G}_{n,k} \sum\limits_{q=1}^\infty [\Sigma_{k,m} \mathcal{G}_{m,n'}]^q  \ . \label{G-Sigma}
 \end{equation}
     The straightforward resummation in (\ref{G-Sigma}) yields the  Green function in the first Born approximation:  
 \begin{equation}   \langle G_{\mathbf{k}} (\omega)\rangle = \left[ \mathcal{G}^{-1}_{\mathbf{k}}(\omega)-\Sigma^{\rm (BA)}_{\mathbf{k}}(\omega)\right]^{-1} \ .
    \end{equation} 
    Here, the self-energy is given by   $ \Sigma_{n,n'}^{\rm (BA)}=\langle \mathcal{V}_{n,k} \mathcal{G}_{k,m}   \mathcal{V}_{m,n'}\rangle$.
   After the averaging one finds that the self-energy is local in the real space and reads:
   \begin{equation}
 \Sigma_{n,n'}^{\rm (BA)}(\omega; u,w)
 =\gamma^2 \delta_{n,n'} \sigma_x \mathcal{G}_{n,n}(\omega; u,w) \sigma_x \ .
 \end{equation} 
 Here, $\mathcal{G}_{n,n}=\int\limits_{-\pi}^\pi\mathcal{G}_{\mathbf{k}}\frac{d{\mathbf{k}}}{2\pi}$. Odd terms in ${\mathbf{k}}$ cancel out under the integration and we have the following structure of the self-energy in $\sigma$-space:
    \begin{equation}
 \Sigma^{\rm (BA)}_{\mathbf{k}}(\omega;u,w)=  \gamma^2\Big[ f(\omega;u,w)\sigma_0  +g(\omega;u,w)\sigma_x \Big]\ . \label{Sigma-BA}
 \end{equation} 
Here,
 \begin{equation}
f(\omega;u,w)=   \int\limits_{-\pi}^\pi  \frac{\omega+i\alpha  }{(\omega+i\alpha)^2 - \epsilon_{\mathbf{k}}^2(u,w)} \frac{d{\mathbf{k}}}{2\pi}  \ . 
\label{int-f}
 \end{equation} 
 and 
  \begin{equation}
g(\omega;u,w)=  \int\limits_{-\pi}^\pi  \frac{ u+w\cos \mathbf{k} }{(\omega+i\alpha)^2 - \epsilon_{\mathbf{k}}^2(u,w)} \frac{d{\mathbf{k}}}{2\pi} \ . \label{int-g}
 \end{equation} 
    We observe that functions $f$ and $g$   renormalize the frequency and hopping element $u$ as follows:      \begin{equation}
    \omega\to \omega-\gamma^2f(\omega;u,w)
      \end{equation}  
      and    
       \begin{equation}
     u\to u+\gamma^2g(\omega;u,w)  \ .
        \end{equation} Finally, 
    we find that the averaged Green function   is represented via  the bare Green function (\ref{bare-GF}) as follows:
     \begin{multline}
  \langle G_{\mathbf{k}} (\omega;u,w,\gamma)\rangle  
  = \\ =\mathcal{G}_{\mathbf{k}}\left(\omega-\gamma^2f(\omega;u,w); u+\gamma^2g(\omega;u,w),w \right)   \ .  \label{G-BA}
 \end{multline} 
The renormalization in (\ref{G-BA}) allows to construct a self-consistent Born approximation procedure. We leave this issue beyond  the scope of our consideration.

\subsection{Band-touching condition} 
\label{band-touching}
We address the limit of weak dimerization, $|u-w|\ll u$, when the gap in the clean limit is small compared to the bandwidth.  The low-energy modes reside close to the momentum $\mathbf{k}=\pm \pi$. We  approximate the spectrum near this point  reads as $\epsilon_{\mathbf{q}}(\Delta,u)=\sqrt{\Delta^2+u^2\mathbf{q}^2}$; we  introduced here the  momentum counted from the edge of the Brillouin zone, $\mathbf{q}=\mathbf{k}-\pi$, and  the  dimerization parameter  $\Delta=u-w$ which is small compared to $u$. We note that an exact calculation of $g$ via the contour integrals with $z=e^{i{\mathbf{k}}}$ gives $g=0$ at $\omega=0$ and $\Delta<0$. Consequently, the band-touching condition is only possible for  $\Delta>0$ which is assumed hereafter. 

Calculation of the integrals (\ref{int-f}) and (\ref{int-g}) is reduced to an integration of a very narrow Lorentian peak near $\mathbf{q}=0$ within the   approximation indicated.  We find that for $\omega=0$, that corresponds to local Green function at the midgap, the integrals  are
 \begin{multline}
f(\Delta, u )\approx -  \int\limits_{-\infty}^\infty  \frac{i\alpha  }{{\mathbf{q}}^2 + \frac{\Delta^2+\alpha^2}{u^2} } \frac{d{\mathbf{q}}}{2\pi u^2}  = \\ =\frac{-i\alpha}{2 u \sqrt{\Delta^2+\alpha^2}}   \label{f-1}
 \end{multline} 
 and 
  \begin{multline}
g(\Delta, u )\approx   -  \int\limits_{-\infty}^\infty  \frac{ \Delta }{{\mathbf{q}}^2 + \frac{\Delta^2+\alpha^2}{u^2} } \frac{d{\mathbf{q}}}{2\pi u^2} = \\ =   \frac{-\Delta}{2 u \sqrt{\Delta^2+\alpha^2}} \ . \label{g-1}
 \end{multline} 
The diagonal component of the Green function at coincident coordinates near  $\omega=0$ reads: 
\begin{equation}\mathcal{G}^{\rm (diag)}= f(\omega-\gamma^2f(\Delta,u);u+\gamma^2g(\Delta,u),w) \ .
\end{equation}
It has the following form
 \begin{multline}
\mathcal{G}^{\rm (diag)}(\omega)\approx\\
\approx -  \int\limits_{-\infty}^\infty  \frac{\omega-\gamma^2f(\Delta,u)+i\alpha  }{{\mathbf{q}}^2 + \frac{(\Delta+\gamma^2 g(\Delta,u))^2-(\omega-\gamma^2f(\Delta,u)+i\alpha)^2}{u^2} } \frac{d{\mathbf{q}}}{2\pi u^2} = \\
= -\frac{\omega-\gamma^2f(\Delta,u)+i\alpha}{2 u \sqrt{(\Delta+\gamma^2 g(\Delta,u))^2+(\omega-\gamma^2f(\Delta,u)+i\alpha )^2}}  \ .
 \end{multline} 
 The  imaginary part of  the Green function   allows to obtain the spectral density of states   as 
 \begin{equation}
  \rho(\omega) =\frac{-1}{\pi}{\rm Im} \mathcal{G}^{\rm (diag)}(\omega) \ . 
 \end{equation} 
 For the midgap energy, i.e., $\omega=0$, one finds:
 \begin{multline}
 \rho(0) = \frac{\alpha(1+\frac{\gamma^2}{2u\Delta})}{2\pi u \sqrt{(\Delta-\frac{\gamma^2 \theta(\Delta)}{2u} )^2+\alpha^2(1+\frac{\gamma^2}{2u\Delta})^2}}  \ . \label{DOS}
 \end{multline}
Here, we use approximate form of $f$ and $g$ from (\ref{f-1}) and (\ref{g-1}). 
 The midgap density of states given by (\ref{DOS}) has a singularity when  the band-touching condition    holds:
 \begin{equation}
 \Delta-\frac{\gamma^2  }{2u}=0 \ . \label{cond-gap}
 \end{equation}
The equation is resolved as $
 \gamma_0^{\rm (BA)} = \sqrt{2u(u-w)}$ where $  u>w>0$. 
In other words, the critical disorder strength found after first Born approximation    reproduces the result (\ref{BA-gamma-0}) derived  via central limiting theorem.

   
%

\end{document}